\documentstyle[aps,preprint]{revtex}
\begin{document}
\draft

\title{Radiative Muon Capture and 
Induced Pseudoscalar Coupling Constant in Nuclear Matter}

\author{Myung Ki Cheoun
\footnote{e.mail : cheoun@phya.yonsei.ac.kr}, 
K.S.Kim, B.S.Han and Il-Tong Cheon}

\address{
Department of Physics, Yonsei University, Seoul, 120-749, Korea
\\ (10 November, 1998)}

\maketitle
\begin{abstract}
The recent TRIUMF experiment for $\mu^- p \rightarrow n \nu_{\mu} 
\gamma$ gave
a surprising result that the induced pseudoscalar coupling constant
$g_P$ was larger than the value obtained from
$\mu^- p \rightarrow n \nu_{\mu}$ experiment as much as 44 \%.
Reexamining contribution of the axial vector current in 
electromagnetic interaction,
we found an additional term to
the matrix element which was used
to extract the $g_P$ value from the measured photon
energy spectrum. This additional term, 
which plays a key role to restore the
reliability of $g_P ( - 0.88 m_{\mu}^2 ) = 6.77 g_A (0)$,
is shown to affect the $g_P$ quenching problems in nucleus.

\end{abstract}
\vspace{1cm}
\pacs{PACS numbers : 25.30.-c, 23.40.Bw, 21.65.+f}

\newpage

\section{Introduction}
The matrix element of vector and axial vector currents
are generally given as
\begin{eqnarray}
\langle N (p^{'}) \vert V_a^{\mu} (0) \vert N (p) \rangle & =
{\bar u} ( p^{'}) [ G_V ( q^2) \gamma^{\mu} + {{G_S ( q^2 )} \over { 2 m}}
q^{\mu} + G_M ( q^2) \sigma^{\mu \nu} q_{\nu} ] {\tau_a \over 2} u(p)
   \nonumber \\
\langle N (p^{'}) \vert A_a^{\mu} (0) \vert N (p) \rangle &=
{\bar u} ( p^{'}) [ G_A ( q^2) \gamma^{\mu} + {{G_P ( q^2 )} \over { 2 m}}
q^{\mu} + G_T ( q^2) \sigma^{\mu \nu} q_{\nu} ] \gamma_5 
{\tau_a \over 2} u(p)~, \nonumber \\
\end{eqnarray}
where $G_A (0) = g_A (0),~ G_M(0) = g_M(0),~ G_V(0) = g_V(0)$ and 
$~G_P ( q^2) = ( {{ 2 m } \over { m_{\mu}}} )
g_P ( q^2) $ with the nucleon and muon masses, $m $ and $ m_{\mu}$. 
$\tau_a$ is the isospin operator. $G_S$ and $G_T$ belong to the
second class current which has a different G-parity from the first
class current, and they are assumed to be absent from the
muon capture to be discussed in this paper. 
On the basis of PCAC (Partially
Conserved Axial Current), the induced pseudoscalar coupling constant is
calculated as
\begin{equation}
g_P ( -0.88 m_{\mu}^2 ) = { { 2 m ~  m_{\mu} } \over 
{ m_{\pi}^2 + 0.88 m_{\mu}^2}} g_A (0) = 6.77 g_A(0 ).
\end{equation}
This value is confirmed by an experiment of the 
ordinary muon capture (OMC) on a proton, ${\mu}^- p \rightarrow
n \nu_{\mu}$ \cite{Ba81}.

However, in order to obtain more precise data, the TRIUMF group 
measured recently the photon energy
spectrum  of the radiative muon capture (RMC) on a proton, $\mu^- p \rightarrow
n \nu_{\mu} \gamma$ and extracted a surprising result \cite{Jo96}
\begin{equation}
{\hat g_P} \equiv g_P ( - 0.88 m_{\mu}^2 ) / g_A (0) = 9.8 \pm 0.7 \pm 0.3~.
\end{equation}
It exceeds the value obtained from OMC as much as 44\%. 
This discrepancy is serious because the theoretical value of
$g_P$ is predicted in a fundamental manner
based on PCAC and agrees with the OMC value.
As long as PCAC is assumed to be creditable, a doubt may be
cast on the result of TRIUMF experiment.
Recent calculations \cite{Me97,An97} by chiral perturbation
also says such a doubt. However, in order to solve this puzzle,
one has to reexamine carefully 
the Beder-Fearing formula \cite{Fe80,Be87}, which is a phenomenological model, 
used to extract the $g_P$ value from the measured RMC spectrum.

In finite nuclei, through the theoretical 
analyses of OMC experimental results,
it is already reported \cite{Ha96} that the ${\hat g}_P$ value is quenched
in medium-heavy and heavy nuclei while it is enhanced in light nuclei.
Since these analyses are carried out before the recent TRIUMF experiment one
needs to reconsider those analyses from another viewpoint.

In this paper we present more successful analysis for the recent TRIUMF
data and show some progressive results for ${\hat g}_P$ quenching problems 
in nucleus by applying our results on proton to nuclear matter.

\section{Basic Formulae}
We start from the ordinary linear-$\sigma$ model ;
\begin{equation}
{\cal L}_0 = {\bar \Psi} [ i \gamma^{\mu} \partial_{\mu} -
g ( \sigma + i {\vec \tau} \cdot {\vec \pi} \gamma_5 )] \Psi
+ { 1 \over 2} [ {( \partial_{\mu} {\vec \pi} )}^2 +
{( \partial_{\mu}{\sigma} )}^2 ]
 + { 1 \over 2} {\mu}^2 ( {\vec \pi}^2 + {\sigma}^2 )
- { {\lambda}^2 \over 4} {( {\vec \pi}^2 + {\sigma}^2 )}^2
\end{equation}
, which gives the following axial current
\begin{equation}
A_{\mu}^a = {\bar \Psi} \gamma_{\mu} \gamma_5 { {\tau_a} \over 2} \Psi
+ {\pi}^{a} {\partial}_{\mu} \sigma - \sigma {\partial}_{\mu} {\pi}^a ~.
\end{equation}
By the spontaneous breakdown of chiral symmetry, $\sigma$ field is shifted
to ${\sigma}^{'} = \sigma - {\sigma}_0$ with ${\sigma}_0 = f_{\pi}$. 
Consequently, the pion appears as Nambu-Goldstone boson. The PCAC can be 
satisfied by the additional inclusion of the explicit chiral symmetry breaking
term as well known.

But the axial current
\begin{equation}
A_{\mu}^a = {\bar \Psi} \gamma_{\mu} \gamma_5 { {\tau_a} \over 2} \Psi
- f_{\pi} {\partial}_{\mu} {\pi}^a ~
\end{equation}
gives $g_A = 1$ in the tree approximation. Following the
recipe of Akhmedov \cite{Akh89} to cure this problem, we add chiral invariant 
lagrangian ${\cal L}_1$ to ${\cal L}_0$,
\begin{equation}
{\cal L}_1 = C[ {\bar \Psi} {\gamma}_{\mu} {{\vec \tau} \over 2} \Psi
( {\vec \pi} \times {\partial}_{\mu} {\vec \pi})
+ {\bar \Psi} {\gamma}_{\mu} \gamma_5 {{\vec \tau} \over 2} \Psi
( {\vec \pi}  {\partial}_{\mu} {\sigma} - \sigma {\partial}_{\mu} {\vec \pi})
]~,
\end{equation}
where arbitrary parameter $C$ is determined so that the 
axial current pertinent to nucleons in ${\cal L} = {\cal L}_0 + {\cal
 L}_1$
\begin{equation}
{}^{(N)} {A_{\mu}^{a}} = {\bar \Psi} {\gamma}_{\mu} {\gamma_5} 
{{{\tau}_a } \over 2} \Psi
[ 1 + C^2 ( {\vec \pi}^2 + {\sigma}^2 ) ]
\end{equation}
should satisfy ${}^{(N)}{A_{\mu}^{a}} = 
g_A {\bar \Psi} {\gamma}_{\mu} {\gamma}_5 {{{\tau}_a} \over 2} \Psi
$ with $g_A = 1.26$. The Goldberger-Treiman relation then 
is satisfied exactly. As a consequence, ${}^{(N)} A_{\mu}$ includes
the contribution not only from the nucleon but also from the 
$\pi - N$ interactions.

Now, the axial vector current consists of the nucleon and pion sectors as
\begin{eqnarray}
A^{\mu}_a ( x) = {}^{(N)}\! A_a^{\mu} ( x) + {}^{(\pi)} A_a^{\mu} (x) 
  \\ \nonumber
= {}^{(N)} A_a^{\mu} ( x) +  f_{\pi} \partial^{\mu}  {\phi}_a (x)~, 
\end{eqnarray}
where $f_{\pi}$ is the pion decay constant. $\phi_a (x) $ is the pion field.

To describe RMC, we need a radiative axial current, which
is used to obtain the transition amplitude of RMC by coupling to
the weak current of lepton line. 
Three different methods are considered 
in order to construct a radiative axial
current. The 1st method \cite{Ch98} is to start from
the above lagrangians ${\cal L} = {\cal L}_0 + {\cal L}_1$ using 
covariant derivative, by which we introduce 
a photon field in U(1) gauge invariant
way. The outcoming lagrangian gives a radiative axial current,
which characteristic is its non-conservation through
the explicit chiral symmetry breaking due to the electromagnetic 
interaction. The second method \cite{Ch97} is to use the extended Euler
equation \cite{Ad65} for the lagrangian 
${\cal L} = {\cal L}_0 + {\cal L}_1$.
The third is to make it directly from the
above axial currents, eq.(9), by exploiting it minimal coupling scheme 
to the momenta of
relevant particles. Here we follow the third one. 
Of course the final radiative axial currents from 
these three different methods 
turned out to be equivalent.

Let us begin from the 
divergences of $^{(N)} A_a^{\mu} (x)$ and $^{(\pi)} A_a^{\mu} (x)$,
\begin{equation}
\partial_{\mu} {}^{(N)} A_a^{\mu} (x) = {\partial}_{\mu} 
[ g_A {\bar \Psi} (x) {\gamma}^{\mu}
\gamma_5 { \tau_a \over 2} \Psi (x) ] \equiv f_{\pi} J_a^{N}~,
\end{equation} 
\begin{equation}
\partial_{\mu} {}^{(\pi)} A_a^{\mu} (x) =  f_{\pi} \partial^2 \phi_a (x) =
- f_{\pi} [ m_{\pi}^2 \phi_a (x) + J_a^N ]~,
\end{equation} 
where we used pion field equation from the above lagrangians 
\begin{equation}
(\partial^2 + m_{\pi}^2 ) \phi_a = -J_a^N~.
\end{equation}
The quantity $J_a^N$ denotes the pion source term. Therefore,
the divergence of total axial currents is given in the following way 
\begin{eqnarray}
\partial_{\mu} A_a^{\mu} (x) & = &
\partial_{\mu} {}^{(N)} A_a^{\mu} (x) + \partial_{\mu} {}^{(\pi)} A_a^{\mu} (x) 
\\ \nonumber
& =& - f_{\pi} [ m_{\pi}^2 \phi_a (x) + J_a^N ] + f_{\pi} J_a^N \\ \nonumber
& =& - f_{\pi} m_{\pi}^2 \phi_a (x) ~.
\end{eqnarray}
This is PCAC.
By eqs. (10) and (12), one can obtain
\begin{equation}
\phi_a (x) = -{ 1 \over { f_{\pi} ( \partial^2 + m_{\pi}^2 )}} 
\partial_{\mu} {}^{(N)}\!A_a^{\mu} (x).
\end{equation}
Substitution of eq.(14) into eq.(9) yields
\begin{equation}
A_a^{\mu} (x) = {}^{(N)}\!A_a^{\mu} (x) - {i \over
{ \partial^2 + m_{\pi}^2} } (i\partial)^{\mu} [ \partial_{\mu} {}^{(N)}A_a^{\mu} (x) ].
\end{equation}
In order to clarify a role of the axial current
in description of the radiation process, we adopt the minimal
coupling prescription, i.e. $\partial_{\lambda} \rightarrow 
\partial_{\lambda} - ie {\cal A}_{\lambda}$ and 
$q^{\mu} \rightarrow 
q^{\mu} - e {\cal A}^{\mu}$. This procedure leads to 
\begin{eqnarray}
A_a^{\mu} (x) & = & {}^{(N)}\! A_a^{\mu} (x) -  
{ i \over {\partial^2 + m_{\pi}^2 }} (i\partial)^{\mu} [
\partial_{\lambda} {}^{(N)} A_a^{\lambda} (x)] 
+ { i e \over {\partial^2 + m_{\pi}^2  }} \epsilon^{\mu} [
\partial_{\lambda} {}^{(N)} A_a^{\lambda} (x)]  \nonumber \\
&&- { e \over {\partial^2 + m_{\pi}^2  }} (i\partial)^{\mu} [
\epsilon_{\lambda} {}^{(N)} A_a^{\lambda} (x)]
+ { e^2 \over {\partial^2 + m_{\pi}^2  }} [
\epsilon^{\mu}
\partial_{\lambda} {}^{(N)} A_a^{\lambda} (x)]~,
\end{eqnarray}
where the potential ${\cal A}_{\lambda}$ is replaced by the photon
polarization vector $\epsilon_{\lambda}$. Notice here that
$\partial_{\lambda} {}^{(N)}A_a^{\lambda} (x) =
g_A m {\bar \Psi} (x) i \gamma_5 \tau_a \Psi (x)$. The last term in eq.(16)
can be neglected because it appears at $O(e^2)$ order. 
Thus we can express the axial current in the
radiative processes in the following realistic form 
\begin{eqnarray}
 A_a^{\mu} (x)
& = & {\bar \Psi} (x) [ g_A \gamma^{\mu} \gamma_5 +
{  {g_P (q^2)  } \over { m_{\mu}}} q^{\mu} \gamma_5
- {  {e g_P ( q^2)  } \over { m_{\mu}}} \epsilon^{\mu} \gamma_5 ]
{\tau_a \over 2} \Psi (x) \\ \nonumber
& & - {  {e g_P (q^2)  } \over { 2 m m_{\mu}}} q^{\mu}
[ {\bar \Psi} (x) \epsilon_{\alpha} \gamma^{\alpha} \gamma_5
{\tau_a \over 2} \Psi (x) ]~.
\end{eqnarray}
The fourth term, which corresponds to ``Seagull term'', is
missing \cite{Ch97} in the previous calculations \cite{Fe80,Be87}. 

Following the Fearing's formulation and notation \cite{Fe80} for the diagrams 
given in ref. \cite{Fe80}, one can evaluate the relativistic amplitude of RMC
on a proton as
\begin{equation}
M_{ f i} =
M_a + M_b + M_c + M_d + M_e + \Delta M_e
\end{equation}
with
\begin{eqnarray}
M_a &=& - \epsilon_{\alpha} {\bar u}_n \Gamma^{\delta} ( Q) u_p
             \cdot {\bar u}_{\nu} \gamma_{\delta} ( 1 - \gamma_5)
  { { \mu\!\!\!/ - {k \!\!\!/} + m_{\mu}  } \over { - 2 k \cdot \mu}} 
\gamma^{\alpha} u_{\mu} ~,\\ \nonumber
M_b &=&  \epsilon_{\alpha} L_{\delta} {\bar u}_n \Gamma^{\delta} ( K)
{  {{p \!\!\!/}-{ k \!\!\!/}+ m_p} \over {- 2 k \cdot p}} 
(\gamma^{\alpha} - i \kappa_p {  \sigma^{\alpha \beta} \over { 2 m_p}} 
k_{\beta} ) u_p ~, \\ \nonumber
M_c &=&  \epsilon_{\alpha} L_{\delta} {\bar u}_n 
( - i \kappa_n { \sigma^{\alpha \beta} \over {2 m_n  }} k_{\beta} )
{  {{n \!\!\!/}+{ k \!\!\!/}+ m_n} \over { 2 k \cdot n}} 
 \Gamma^{\delta} ( K)  u_p ~,\\ \nonumber
M_d &=&  - \epsilon_{\alpha} L_{\delta} {\bar u}_n 
( { {2 Q^{\alpha}  + k^{\alpha}} \over { Q^2 - m_{\pi}^2 }} 
{ g_P ( K^2)  \over m_{\mu} }
  K^{\delta} \gamma_5 ) u_p ~,\\ \nonumber
M_e &=&   \epsilon_{\alpha} L_{\delta} {\bar u}_n 
( {   {i g_M } \over {2 m}  } \sigma^{\delta \alpha}
 + {{g_P ( Q^2)}  \over{m_{\mu}} } 
\gamma_5 g^{\delta \alpha} )  
 u_p ~,\\ \nonumber
\Delta M_e & =&  
- \epsilon_{\alpha} L_{\delta} {\bar u}_n 
( {   { g_P (K^2)} \over {2 m  m_{\mu}}  }  Q^{\delta} \gamma_5 
\gamma^{\alpha} ) u_p ~,
\end{eqnarray}
where 
\begin{equation}
\Gamma^{\delta} ( q ) = g_V \gamma^{\delta} + { { i g_M } \over { 2 m}}
\sigma^{\delta \beta } q_{\beta} + g_A \gamma^{\delta} \gamma_5 +
{  {g_P (q^2)} \over m_{\mu}} q^{\delta} \gamma_5 ~,
\end{equation}
$L_{\delta} = {\bar u}_{\nu} \gamma_{\delta} ( 1 - \gamma_5) u_{\mu},
K = n - p + k $ and $ Q = n - p$ with momenta of neutron,
proton and photon, $ n, p $ and $ k$, respectively. And $m \sim m_p 
\sim m_n$. Other constants are taken as $g_V = 1.0, g_A = - 1.25, g_M = 3.71,
\kappa_p = 1.79$ and $\kappa_n = - 1.91$ \cite{Fe80}. $M_e$ term is
originated from the third term in eq.(17) and 
$\Delta M_e$ term comes from the fourth term. 
But the latter, $\Delta M_e$, is missing in the paper by Fearing
\cite{Fe80,Be87}. Accordingly, this term was not included in
the previous procedure of extracting $g_P$ value from the experimental 
RMC photon energy spectrum \cite{Jo96}. 

The above transition amplitude can be also understood in terms of 
pseudo vector (PV)
coupling scheme between nucleons and virtual pion, through which
external axial current interacts with nucleons. Moreover
in nuclear matter, this PV coupling type is preferred rather than PS
coupling type because the former is consistent with
PCAC while the latter  contradicts to PCAC in nuclear matter \cite
{Akh89}.

The RMC transition rate is given by
\begin{equation}
{ {d \Gamma_{RMC}} \over {d k}} =
{  {\alpha G^2 \vert \phi_{\mu} \vert^2 m_N  } \over {  {( 2 \pi )}^2 }}
 \int_{-1}^{1}  dy 
{  { k E_{\nu}^2  } \over { W_0 - k( 1 - y) }} { 1 \over 4} \sum_{spins}
 \vert M_{f i } {\vert}^{2} ~,
\end{equation}
where $\alpha$ is the fine structure constant, $G$ 
is the standard weak coupling constant, $ y = {\hat k } \cdot 
{\hat \nu},~ k_{max} = ( W_0^2 - m_n^2 ) / 2 W_0,~ E_{\nu} = W_0
( k_{max} - k ) / [ W_0 - k( 1 - \nu)],~ W_0 = m_p + m_n $ -
(muon binding energy) and $\vert \phi_{\mu} \vert^2$ is the absolute square
of muon wave function averaged over the proton which
is taken as a point Coulomb. 
In order to compare to the experimental results, we take the following 
steps. For liquid hydrogen target, muon capture is
dominated through the ortho and para $p \mu  p$ molecular states
\cite{Jo96,Ba82}. Since these molecular states can be attributed to the
combinations of hyperfine states of $\mu  p$ atomic states \cite{Ba82}
i.e. single and triplet states, we decompose the statistical spin
mixture ${1 \over 4} \sum_{spins} \vert M_{ f i } {\vert}^2$ into
such hyperfine states by reducing $4 \times 4 $ matrix elements to
$2 \times 2$ spin matrix elements. At this step, we
confirmed that when the $\Delta M_e$ term was not
included, eq.(21) reproduced the curves given in ref. \cite{Be87}. 

For the description of RMC in nuclear matter, we follow the Fearing's 
paper \cite{Fe89}, i.e. we adopted the relativistic
mean field theory \cite{Se86} where the nucleons are treated as free Dirac
particles with effective mass due to the scalar and vector potential. Then
the nucleus are the Fermi gas. Our RMC capture rate in nuclear matter 
is given in the following way 

\begin{eqnarray}
\Gamma_{RMC}^{NM}&  = &{{\alpha G^2 } \over { 4 {\pi}^2}} 
{  { 3 m_P m_N } \over { 4 \pi k_F^3}} \vert \phi_{\mu} {\vert}^2 
\int_{k_{min}}^{k_{max}} dk 
\int_{{(cos {\theta}_k )}_{min}}
^{{(cos {\theta}_k )}_{min}} d cos {\theta}_k \int_{p_{min}}^{k_F} dp
\int_{k_F}^{n_{max}} dn  \nonumber \\ 
& & \int_{0}^{ 2 \pi} d \phi_n 
{  {  E_{\nu} n p^2 k } \over { E_p E_n \vert {\vec n} + {\vec p} \vert}}
{1 \over 4} \sum_{spins}
 \vert M_{f i } {\vert}^{2} ~,
\end{eqnarray}
where the integration intervals of nucleon momenta, which comes from 
the Pauli blocking in nuclear matter, are calculated in detailed kinematics 
and $k_F$ is the Fermi momentum. For finite nuclei, we need to know the 
corresponding muon wave functions and 
$k_F$ values, but which depends on the model. We already suggested a model
\cite{Ch97} for this purpose, but will be skipped here and concentrate on the 
case of nuclear matter. 

\section{Results and Discussions}

Our results for RMC on proton are shown in Fig.1. 
The solid curve is the spectrum obtained
in ref. 2., i.e. the result without $\Delta M_e$ term for 
${\hat g}_P = 9.8$. On the other hand, the dotted curve is
calculated without $\Delta M_e$ term for ${\hat g}_P$ = 6.77. This curve 
is obviously much lower than the solid curve. When 
$\Delta M_e$ term is taken into account for ${\hat g}_P$ = 6.77, we obtain
the dashed curve which is very close to the solid curve for
the energy spectrum on $k \ge 60 MeV$. 
The minor discrepancy
may be due to the neglect of higher order contributions and
other degree of freedom such as $\Delta$. Our result shows that $\Delta M_e$
term restores the credit of ${\hat g}_P = 6.77$.

The number of RMC photons observed for $k \ge 60 MeV $ is 279 $\pm$ 26 and
the number of those from the solid curve is 299, while 
our result obtained by integrating the dotted curve
spectrum is 273. Since the contribution of $\Delta$ degree of freedom
is known to be a few percent \cite{Be87}, it is not included in the present
calculation. Vector mesons such as $\rho$ and $\omega$ are also turned out
to have very
small contributions in this calculation. Higher order terms are
pointed out to be insignificant \cite{Fe80}.

It is confirmed that the matrix elements upto the $M_e$ term in eq.(19) satisfy
gauge invariance. However, it is broken when the $\Delta M_e$ term and
$\Delta$ degree of freedom are included. As far as diagram method
is adopted, the gauge invariance is more or less broken because
a series of diagrams will be cut somewhere. In order to estimate 
the rate of gauge invariance 
broken by the $\Delta M_e$ term, we evaluated the spectrum with $\Delta M_e$
alone. The result is shown by a dot-dashed curve in Fig.1. The
gauge invariance breaking is not so large as expected. 
However, in the limit $m_{\mu} \rightarrow 0$, the gauge invariance
is restored.

In spite of such a burden of gauge invariance,
our present calculation shows that ${\hat g}_P$ = 6.77 is
reasonable for both OMC and RMC on a proton.

Figure 2 shows the photon energy spectrum in nuclear matter, which is 
presented as 
the ratio of RMC to OMC in order to reduce the uncertainty from the nuclear
structure. We compared our amplitude to Fearing's analysis, which is PS 
coupling and has been used
to extract the ${\hat g}_P$ from the experimental data.
At the same ${\hat g}_P$, our ratio R is higher than the PS coupling scheme.
It means that ${\hat g}_P$ value to fit some experimental data
becomes lower in our PV scheme. Therefore ${\hat g}_P$ quenching rate 
is smaller than the usual PS coupling scheme.
These behaviour are nearly independent of effective nucleon mass in
nuclear matter (see the solid, dotted and dashed lines).

Figure 3 shows another interesting results for ${\hat g}_P$ quenching
in finite nuclei. The larger $k_F$, which may mean the heavier nuclei, 
the smaller becomes the ratio R. As a result, ${\hat g}_P$ quenching may be 
larger in heavier nuclei. 
But in relatively lower $k_F$ region just reversed results are shown.
Consequently ${\hat g}_P$ may be enhanced in lower $k_F$ region.
Since we have to integrate the nucleon's Fermi motion 
from the possible lowest momentum up to the Fermi momentum, 
the mechanism in lower $k_F$ region 
could play a role of compensating ${\hat g}_P$ quenching 
in larger $k_F$ region. Therefore this could 
be an indication for the
${\hat g}_P$ enhancement in light nuclei.

\section*{Acknowledgments}

This work was supported by the KOSEF and the Korean Ministry
of Education (BSRI-97-2425).

\newpage
\centerline{Figure Captions}

Figure 1. Photon energy spectrum for triplet states in RMC on
proton. The solid curve is deduced without $\Delta M_e$ term for ${\hat g}_P$
= 9.8, whose result corresponds to the experimental results in ref.2. 
The dotted curve is 
obtained without $\Delta M_e$ term for ${\hat g}_P = 6.77$. 
The dashed 
curve is with $\Delta M_e$ for ${\hat g}_P$ = 6.77. 
The dot-dashed curve
is calculated with $\Delta M_e$ term alone for ${\hat g}_P$ = 6.77.

\vskip1cm

Figure 2. The photon energy spectrum for the 
ratio of RMC and OMC in nuclear matter. The thick curves are results from
our transition amplitudes, but thin curves are Fearing's amplitudes.
The solid curves are for $ M^* = M $, the dotted curves are for $M^* = 0.57 M$
and the dashed curves are for $M^* = 0.7 M$.

\vskip1cm

Figure 3. The ratio of RMC and OMC versus fermi momentum $k_F$. 
The dot-dashed curve 
($k_{\gamma}$ = 60 MeV) and dotted curve ($k_{\gamma}$ = 80 MeV) 
come from our amplitude, while the solid and long dashed curves are
from Fearing's, respectively from $k_{\gamma}$ = 60 and 80 MeV.

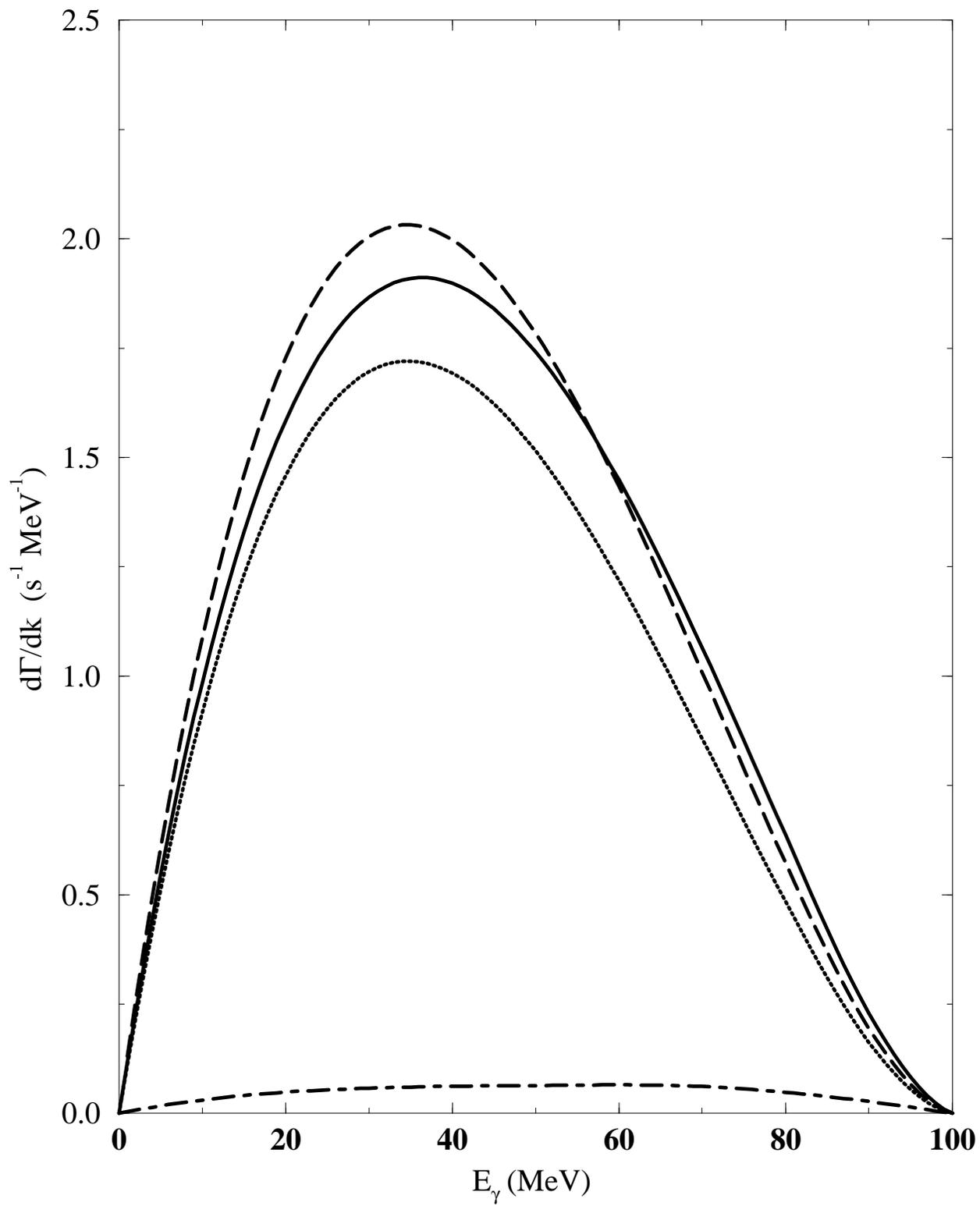
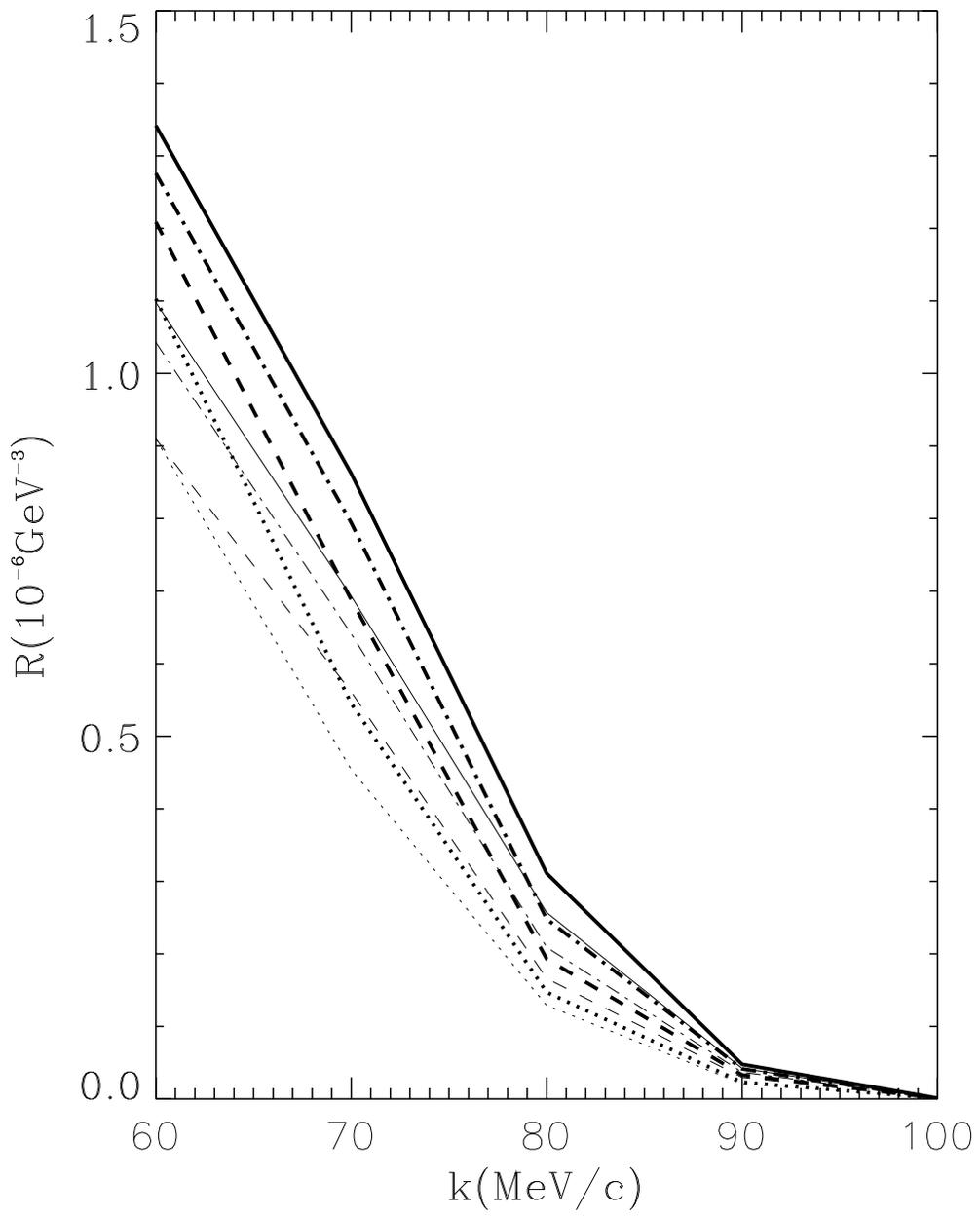
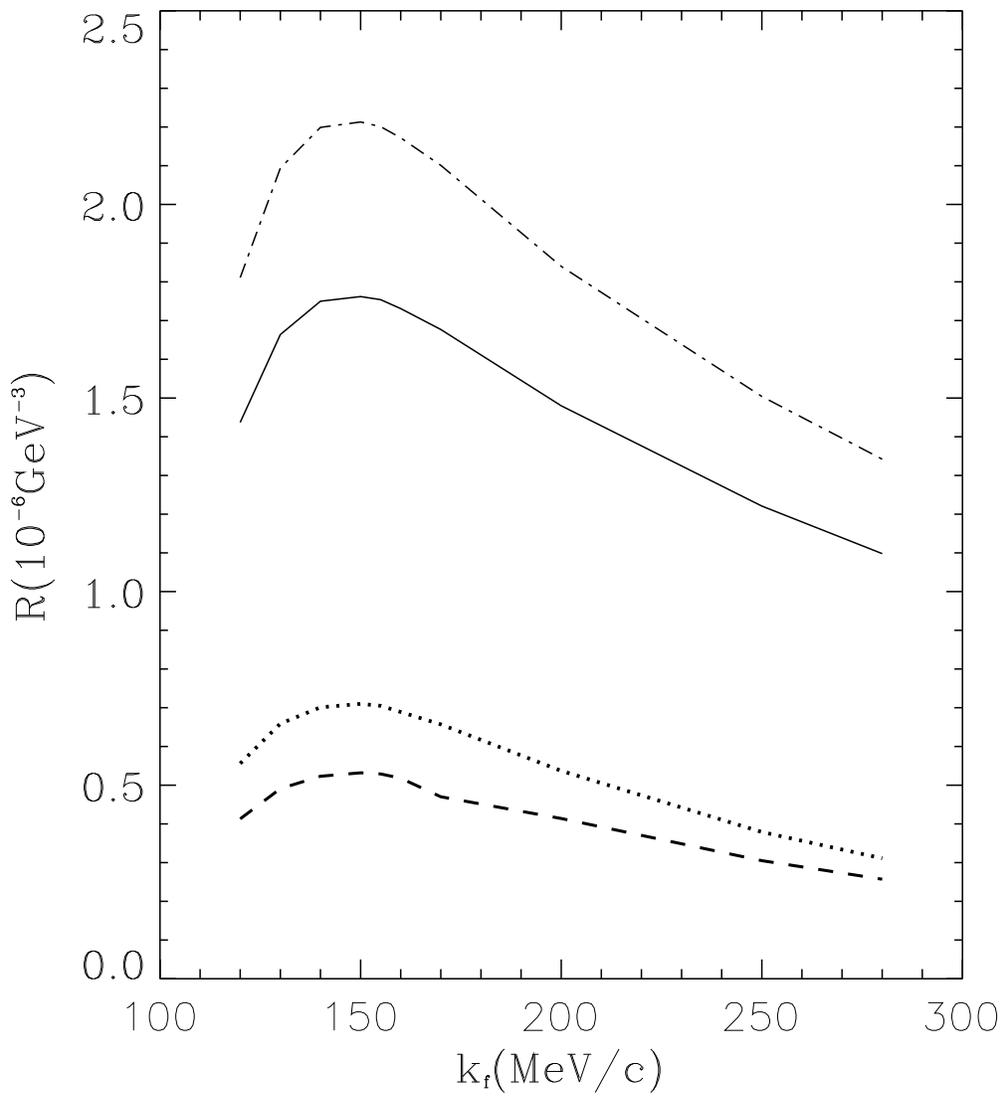
\end{document}